\documentstyle[11pt,aaspp4]{article}

\lefthead{Burgasser et al.}
\righthead{Discovery of 2MASS T Dwarfs}

\begin{document}

\title{Discovery of Four Field Methane (T-type) Dwarfs with the Two Micron All-Sky Survey\footnote{Portions
of the data presented herein were obtained at the W.\ M.\ Keck Observatory
which is operated as a scientific partnership among the California Institute of
Technology, the University of California, and the National Aeronautics and Space
Administration.  The Observatory was made possible by generous financial 
support of the W.\ M.\ Keck Foundation.}}

\author{Adam J. Burgasser\altaffilmark{2},
J. Davy Kirkpatrick\altaffilmark{3},
Michael E. Brown\altaffilmark{4,5},
I. Neill Reid\altaffilmark{6},
John E. Gizis\altaffilmark{7}, Conard C. Dahn\altaffilmark{8},
David G. Monet\altaffilmark{8},
Charles A. Beichman\altaffilmark{9}, James Liebert\altaffilmark{10},
Roc M. Cutri\altaffilmark{3}, and Michael F. Skrutskie\altaffilmark{7}} 

\altaffiltext{2}{Division of Physics, M/S 103-33, 
California Institute of Technology, Pasadena, CA 91125; diver@cco.caltech.edu}
\altaffiltext{3}{Infrared Processing and Analysis Center, M/S 100-22, 
California Institute of Technology, Pasadena, CA 91125}
\altaffiltext{4}{Division of Geological and Planetary Sciences, M/S 105-21, 
California Institute of Technology, Pasadena, California 91125}
\altaffiltext{5}{Alfred P.\ Sloan Research Fellow}
\altaffiltext{6}{Department of Physics and Astronomy, 
University of Pennsylvania, 209 South 33rd Street, Philadelphia, PA 19104-6396}
\altaffiltext{7}{Five College Astronomy Department, Department of Physics
and Astronomy, University of Massachusetts, Amherst, MA 01003}
\altaffiltext{8}{U.S. Naval Observatory, P.O. Box 1149, 
Flagstaff, AZ 86002}
\altaffiltext{9}{Jet Propulsion Laboratory, M/S 180-703, 
4800 Oak Grove Dr., Pasadena, CA, 91109}
\altaffiltext{10}{Steward Observatory, University of Arizona,
Tuscon, AZ 85721}

\begin{abstract}

We report the discovery of four field methane (``T''-type) brown dwarfs using 
Two Micron All-Sky Survey (2MASS) data.  One additional methane dwarf,
previously discovered by the Sloan Digital Sky Survey,
was also identified.
Near-infrared spectra clearly show the 1.6 and 2.2 ${\mu}m$ CH$_4$ absorption bands characteristic
of objects with T$_{eff} {\lesssim}$ 1300 K, as well as broadened H$_2$O bands at
1.4 and 1.9 ${\mu}m$.  Comparing
the spectra of these objects with that of Gl 229B, we propose that all new 2MASS T dwarfs
are warmer than 950 K, in order from warmest to coolest:
2MASS J1217-03, J1225-27, J1047+21 and 
J1237+65.  Based
on this preliminary sample, we find a warm T dwarf surface density of 
0.0022 T dwarfs/sq.\ deg., or
$\sim$ 90 warm T dwarfs over the whole sky detectable to J $<$ 16.
The resulting
space density upper limit, 0.01 T dwarfs/pc$^3$, is comparable to that of the 
first L dwarf sample from Kirkpatrick
et al.

\end{abstract}


\section{Introduction}

Searches for brown dwarfs are meeting with increasing success
in recent years.
Proper-motion surveys (\markcite{Ru97}Ruiz et al.\ 1997), surveys of young 
clusters 
(\markcite{St89}Stauffer et al.\ 1989;
\markcite{Re95}Rebolo, Zapatero-Osorio, \& Mart{\'{\i}}n 1995), 
companion searches (\markcite{Na95}Nakajima et al.\ 1995), 
radial velocity measurements 
(\markcite{My95}Mayor \& Queloz 1995; 
\markcite{Mc95}Marcy \& Butler 1995), and all-sky near-infrared and deep optical surveys 
(\markcite{Df97}Delfosse et al.\ 1997; \markcite{Ki99}Kirkpatrick et al.\ 1999, 
hereafter K99) have identified an 
ever-growing number of confirmed substellar objects.  Until recently, Gl 229B 
(\markcite{Na95}Nakajima et al.\ 1995) stood out amongst those objects as the
only known brown dwarf sufficiently cool (T$_{eff}$ = 960$\pm$70 K, 
\markcite{Ma96}Marley et al.\ 1996) to show CH$_4$ absorption bands at
1.6 and 2.2 $\micron$.
Analysis of preliminary data from the Sloan Digital Sky Survey (SDSS; 
\markcite{Yk99}York et al.\ 1999) has led to the discovery of the first
field counterparts of Gl 229B (\markcite{Ss99}Strauss et al.\ 1999;
\markcite{Ts}Tsvetanov et al.\ 1999).  

In this letter, we report the discovery of an additional four spectroscopically-similar
dwarfs, which we designate as spectral class ``T''\footnote{The primary 
defining feature of spectral class T is the
appearance of the 1.6 and 2.2 micron overtone bands of CH$_4$. While these
objects have also been called ``methane'' dwarfs, we note that the 3.3 micron
fundamental CH$_4$ band is predicted to appear at higher temperatures, i.e.
perhaps amongst the latest-type L dwarfs.} (\markcite{Ki99}K99), based
on data from the Two Micron All-Sky Survey (2MASS; \markcite{Sk97}Skrutskie 
et al.\ 1997); and the recovery of a T dwarf identified by
SDSS.  A brief summary of the 
selection process and initial follow-up of 2MASS T candidates is discussed in 
$\S$2; near-infrared spectroscopy of 
the confirmed T dwarfs
 is discussed in $\S$3; 
and a preliminary sequence and estimates of T$_{eff}$ and the T dwarf space density
are discussed in $\S$4.  Results are summarized in $\S$5.

\section{Candidate Selection}

\subsection{Search Criteria}

Candidates were culled from two separate 2MASS data samples.  The first (sample A), was
taken from the 2MASS Spring 1999 Data Release,\footnote{See the Explanatory 
Supplement to the 2MASS Spring 1999 Incremental Data Release by 
\markcite{Cu99}R. M. Cutri et al., which is at 
http://www.ipac.caltech.edu/2mass/releases/spr99/doc/explsup.html.} 
2483 deg$^2$ of 
northern hemisphere data containing approximately 20.2 million point sources.  Candidates
were constrained to have detections at J and H bands with J $<$ 16 (2MASS 
signal-to-noise ratio $\sim$ 10 limit),
J-H $<$ 0.3 and H-K$_s$ $<$ 0.3,
${\mid}b{\mid} > 15^o$
(to eliminate source confusion in the plane), no minor planet correlations, and 
no optical
counterparts (USNO-A: \markcite{Mo98}Monet et al.\ 1998) within 5$\arcsec$.  The 
adopted near-infrared colors select
objects similar to or cooler than Gl 229B, while excluding the 
overwhelming numbers
of main sequence stars.  Note that this 
may exclude L/T transition objects (0.6 $\leq$ J-K$_s$ $\leq$ 2.1),
although the temperature range over which CH$_4$ absorption causes this transition
may be quite small (\markcite{Rd99}Reid et al.\ 1999).  J- and H-band detections were 
required to exclude false 
sources, although a K$_s$-band detection was not required because of decreased
sensitivity at this wavelength\footnote{K$_s$ non-detections are reported as 
limiting magnitudes (no object detected at given
coordinates to 95\% confidence level).  In this case, the H-K$_s$ color
limit is still required to be less than 0.3 magnitude.}
(2MASS 99.9\% completeness limit is K$_s \sim$ 14.3).  Non-detections 
on POSS-I (\markcite{Mk63}Minkowski \& Abell 1963)
are imposed because these
objects have extremely red optical-infrared colors (Gl 229B R-J $\approx$ 9, 
\markcite{Go98}Golimowski et al.\ 1998).

The second set (sample B) was taken from
3420 deg$^2$  
($\approx$ 28 million point sources) of 
northern and southern hemisphere data, with 
search criteria of J-H $<$ 0.2 and H-K$_s$ $<$ 0.2,
J $<$ 16, ${\mid}b^{II}{\mid} > 20^o$, no minor planet correlations, 
and no POSS-I or 
POSS-II (\markcite{Rd99}Reid et al.\ 1991) detections.  Again, sources were required
to have J- and H-band detections but not required to have K$_s$-band detections.

The search criteria selected 349 candidates from sample A and
319 candidates from sample B.
Subsequent inspection of POSS and 
2MASS data to rule out faint optical sources (background stars)
and high proper motion stars 
eliminated all but 12 candidates from sample A and all but 11 from sample B.
Three candidates from sample A and five from sample B were accessible for
observation in May, 
and are listed in Table 1, which gives the object name (col. [1]),
sample (col. [2]), and 2MASS J-, H-, and K$_s$-band photometry (cols. [3]-[5]).  
The two samples are spatially distinct, and can be considered complete in the 
region of
sky observable during our investigation, 548 deg$^2$ in sample A and 1236
deg$^2$ in sample B.

One additional object,
2MASS J1225-27, which was previously identified as 
a T dwarf candidate (\markcite{Bg98}Burgasser et al.\ 1998), but is not a
member of either sample, 
was also included in our follow-up and is listed in 
Table 1.


\subsection{Imaging Data}

The largest source of contamination among our T candidates is uncatalogued
minor planets (\markcite{Bg98}Burgasser et al.\ 1998; \markcite{Ki99}K99), which
have blue near-infrared colors (J-K$_s \sim$ 0.3, 
\markcite{Ve95}Veeder et al.\ 1995; \markcite{Sy99}Sykes et al.\ 1999) 
and will not match a POSS-I or -II source
because of motion.  While catalogued minor planets are flagged by
the 2MASS processing pipeline, uncatalogued minor planets remain, and
re-imaging is required to 
eliminate them from the candidate pool.  

Eighty-minute z$\arcmin$ band 
(\markcite{Fu97}Fukugita et al.\ 1997)
exposures were 
obtained for two
candidates, 2MASS J1217-03 and 2MASS J1237+65, using the USNO Flagstaff 1.55m Tek2k Camera
on 1999 May 19 (UT).  
The remaining targets in Table 1 were imaged at K-band using 
the Keck I Near 
Infrared Camera (NIRC; \markcite{Mt94}Matthews \& Soifer 1994) on 1999 May 27 (UT).
Results from
these reimaging campaigns are summarized in Table 1, columns (6)-(8).
In total, five of nine objects were confirmed at their expected positions.  
2MASS J-band images of the new confirmed candidates 
are shown in Figure 1.  Designated names 
(col. [1]), 2MASS J magnitudes and J-H, H-K$_s$, and J-K$_s$ colors 
(cols. [2]-[5]), and estimated distances (col. [6]) 
are listed in Table 2.



\section{Spectroscopic Data}

An initial identifying optical (8000 - 11000 {\AA}) spectrum for 2MASS J1237+65 
was obtained from the Palomar
200'' Double Spectrograph (\markcite{Ok82}Oke \& Gunn 1982) on 1999 May 24 
(\markcite{Bg99}Burgasser et al.\ 1999).  The data
identified this object as very cool, as it is similar in appearance to the optical
spectrum of Gl 229B (\markcite{Op98}Oppenheimer et al.\ 1998), though quite
noisy due to the exceedingly faint signal. 
 
All confirmed candidates were then spectroscopically observed in the near-infrared
using the Keck I NIRC 
grism on 1999 May 27-28 (UT).  A 120 line/mm grism with the HK
order sorting filter was employed, for first order resolution 
${\lambda}/{\Delta}{\lambda} = 100$; wavelength resolution was 60 {\AA}/pixel.  
Each target was imaged with NIRC in camera mode and then placed into 
either a 0${\farcs}$52 
(for 2MASS J1217-03, J1225-27, and J1237+65) or 0${\farcs}$38 
(for 2MASS J1047+21 and J1346-00) wide 
slit.  Total integration times of 1000s were divided into
five 200s exposures with 5$\arcsec$ on-slit dithers between exposures. 
The spectra were then pairwise subtracted to remove sky background, and each
spectrum extracted separately.  Flat-fielding was  
performed by obtaining a spectrum of a diffusely illuminated dome
spot with identical instrumental settings.  Each target spectrum was 
divided by a spectrum of a nearby SAO F star to correct for telluric absorption, 
and photometric standards were observed for flux calibrations.  Individual
exposures for each object were then averaged into a single, final spectrum.  

Reduced spectra for all new 2MASS T dwarfs are shown in Figure 2, along with 
Gl 229B\footnote{The NIRC spectrum of the SDSS rediscovery, 2MASS J1346-00,
will be presented in \markcite{Bg99}Burgasser et al.\ (1999).}
(\markcite{Le99}Leggett et al.\ 1999). 
All but one of the objects, 2MASS J1225-27, were 
saturated at the H-band peak, resulting in a false ``absorption'' feature
at 1.58 $\micron$.  The region of saturation was at most 3 pixels wide.  
An extrapolation of each saturated H-band peak using the 
2MASS J1225-27
spectrum is indicated by a dotted line; each spectrum
is normalized to this revised H-band peak.    
The 1.6 and 2.2 ${\mu}m$ methane bands are quite obvious in all of the confirmed
candidates, identifying all as T dwarfs.  Broadened H$_2$O absorption bands
shortward of 1.5 and 2.05 ${\mu}m$ are also evident, consistent with those found
in Gl 299B (\markcite{Op95}Oppenheimer et al./ 1995).
The 2.3 ${\mu}m$ CO band seen in
the L dwarfs (\markcite{Ki99}K99) is either overwhelmed by CH$_4$ absorption 
or absent in the T dwarfs.  

The spectra show significant variation in the ratio
between the flux at 2.1 and 1.55 $\micron$.  As 
Figure 2 indicates, the 2.1 $\micron$ peak flattens from 2MASS J1217-03 down to 
2MASS J1237+65.  In contrast, there is an apparent steepening in the core of the 1.6 $\micron$ CH$_4$
band, which is flat for 2MASS J1217-03, but decreases
from 1.6 to 1.8 $\micron$ for 2MASS J1225-27, J1047+21, and J1237+65.  


\section{Discussion}

\subsection{A Preliminary Sequence and T$_{eff}$}

According to \markcite{Bu97}Burrows et al.\ (1997), the dominant form of
molecular carbon switches 
from CO to CH$_4$ at T$_{eff} \sim$ 1300 K.  H$_2$O vapor bands, 
which have a profound influence on the near-infrared spectra of M 
(\markcite{Jo94}Jones et 
al.\ 1994) and L (\markcite{De97}Delfosse et al.\ 1997) dwarfs, are also expected 
to be 
strong in T dwarfs.  As Tokunaga \& Kobayashi (1999) have shown, H$_2$
collision-induced absorption (CIA) has a marked influence on the 
latest L dwarfs, noticeably
depressing the spectrum in a broad region around 2.2 $\micron$.  This has also
been seen in the reflectance spectrum of Jupiter (\markcite{Da66}Danielson 1966).  
For the lower T$_{eff}$'s of T 
dwarfs, H$_2$ is expected to play a more dominant role.  

Thus, for T$_{eff} \lesssim$ 1300 K, the major absorbers in the
near-infrared (JHK) spectra are CH$_4$, H$_2$O, and H$_2$. 
Since CIA affects the K band more strongly than the H band, particularly toward
lower temperatures,   
the ratio of the flux at
H to that at K should increase with cooler effective temperatures.
This reasoning is the basis of our preliminary spectral sequence as displayed in 
Figure 2: 2MASS J1217-03, 
J1225-27, J1047+21,
J1237+65, and Gl 229B.  The J band is even less influenced by CIA and little 
influenced by CH$_4$, so that 
the same argument should also hold true for 
J-K$_s$; that is, we expect that cooler T dwarfs also have bluer J-K$_s$ colors.
Four of the five objects were not detected at
K$_s$ by 2MASS, however, so these color relations 
require further observational investigation.  

If we assume that these dwarfs have 8.5 $<$ log (age, Gyr) $<$ 9.5,
typical of isolated field objects, then based on the observed colors and the
assumption that all of the objects are warmer than Gl 229B,
the 2MASS T dwarfs are likely to have 
950 $<$ T$_{eff}$ $<$ 1300 K (see Figure 9 of \markcite{Bu97}Burrows et al.\ 1997).  
It must be stressed that this is a preliminary 
assessment of the temperatures of these objects, and no consideration for
other spectral influences such as metallicity and gravity have been made.  
Better T$_{eff}$ estimates of these T dwarfs will require
model fitting of their spectra from the far optical through the near-infrared.
This analysis will be left to a future paper.

\subsection{Space Density and the Brown Dwarf Mass Function}

As indicated in Table 2, objects appear to lie within the J-H $<$ 0.2 and 
H-K$_s <$ 0.2 color region specified for Sample B, even 2MASS J1047+21, which was selected
from a slightly broader color cut.  If we assume that these colors are typical
for all methane dwarfs, then we can make a preliminary
assessment of the warm (900 $<$ T$_{eff} <$ 1300 K) T dwarf space density. 
With one detection in 548 deg$^2$ of sample A and three detections in 
1236 deg$^2$ of sample B, we obtain a mean surface density of
0.0022 T dwarfs deg$^{-2}$, or 90 warm T dwarfs observable over the entire
sky with J $<$ 16.  If we make the simplifying assumption that all four confirmed T dwarfs 
in our sample have 
the same luminosity as Gl 229B, then this J-band limit corresponds to a 
distance limit of ${\sim}$ 13 pc, and thus a 
space density of $\approx$ 0.01 warm T dwarfs pc$^{-3}$.  This value should be
considered an upper limit, as the warmer T dwarfs likely sample beyond 13 pc.  
It is nonetheless comparable
to the L dwarf density of 0.007 L dwarfs pc$^{-3}$ computed by K99.  Comparison
with the simulations of \markcite{Rd99}Reid et al.\ (1999) suggests a mass function 
$\frac{dN}{dM} \propto M^{-1}$ in the T dwarf regime, comparable
to the relation for local late M dwarf stars (\markcite{Rd99}Reid et al.\ 1999).

\section{Summary}

We have identified four new T dwarfs
and one previously discovered T dwarf in 1784 deg$^2$ of 2MASS survey data.
These objects all show the hallmark 1.6 and 2.2 $\micron$ CH$_4$ absorption bands that are
characteristic of objects with T$_{eff} {\lesssim}$ 1300 K.  They show some
variation in their H- to K-band flux ratios, likely due to the combined absorption of CH$_4$,
H$_2$O, and H$_2$.  This allows us to make a preliminary attempt at a 
T dwarf spectral sequence.
We determine a proemial space density estimate of $\lesssim$ 0.01 warm T dwarfs pc$^{-3}$,
comparable to the L dwarf density from K99.

\acknowledgements
A.\ J.\ B.\ would like to thank Tom Geballe and Sandy Leggett for the
use of their recalibrated Gl 229B UKIRT spectrum, Ben Oppenheimer for his
valuable comments, Albert Burgasser for 
consultation on a
T dwarf search database, and especially the 2MASS staff and scientists for their
support and for pointing their
telescopes in the right directions.
J.\ D.\ K.\ acknowledges Michael Strauss, Jill Knapp, and Xiaohui Fan for sharing
news of their T dwarf discovery prior to publication.
A.\ J.\ B.\, J.\ D.\ K.\, I.\ N.\ R.\, and J.\ L.\ acknowledge funding through a NASA/JPL grant to 2MASS
Core Project science. 
A.\ J.\ B.\, J.\ D.\ K.\, R.\ M.\ C.\, and C.\ A.\ B.\ acknowledge the support of the Jet Propulsion
Laboratory, California Institute of Technology, which is operated under
contract with the National Aeronautics and Space Administration.
This publication makes use of data from
the Two Micron All Sky Survey, which is a joint project of the University
of Massachusetts and the Infrared Processing and Analysis Center, funded
by the National Aeronautics and Space Administration and the National
Science Foundation.

\clearpage

\begin{deluxetable}{lccccccc}
\scriptsize
\tablenum{1}
\tablewidth{0pt}
\tablecaption{2MASS T Dwarf Candidates: Imaging Results. \label{tbl-1}}

\tablehead{
\colhead{Object\tablenotemark{a}} &
\colhead{Sample} &
\colhead{J} &
\colhead{H} &
\colhead{K$_s$} & 
\colhead{Date (UT)} & 
\colhead{Instrument} & 
\colhead{Confirmed} \nl
\colhead{(1)} & 
\colhead{(2)} & 
\colhead{(3)} & 
\colhead{(4)} & 
\colhead{(5)} & 
\colhead{(6)} & 
\colhead{(7)} & 
\colhead{(8)}
}
\startdata
2MASSI J1007406$+$180601 & A & $15.75\pm0.06$ & $15.70\pm0.14$ & $15.77\pm0.24$	& 27 May 1999 & Keck NIRC &no\tablenotemark{b}\nl
2MASSI J1047539$+$212423 & A & $15.82\pm0.06$ & $15.79\pm0.12$ & $> 16.29$\tablenotemark{c} & 27 May 1999 & Keck NIRC & yes\nl
2MASSI J1059440$+$183442 & A & $15.95\pm0.07$ & $15.83\pm0.15$ & $15.65\pm0.19$ & 27 May 1999 & Keck NIRC & no\tablenotemark{b}\nl
2MASSW J1217111$-$031113 & B & $15.85\pm0.07$ & $15.79\pm0.12$ & $> 15.91$\tablenotemark{c} & 19 May 1999 & USNO Tek2k & yes\nl  
2MASSW J1225543$-$273947 & --- & $15.23\pm0.05$ & $15.10\pm0.08$ & $15.06\pm0.15$ & 27 May 1999 & Keck NIRC & yes\nl
2MASSW J1237392$+$652615 & B & $15.90\pm0.06$ & $15.87\pm0.13$ & $> 15.90$\tablenotemark{c} & 19 May 1999 & USNO Tek2k & yes\nl
2MASSW J1346464$-$003150\tablenotemark{d} & B & $15.86\pm0.08$ & $16.05\pm0.21$ & $> 15.75$\tablenotemark{c} & 27 May 1999 & Keck NIRC & yes\nl
2MASSW J2110230$-$285509 & B & $15.43\pm0.05$ & $15.31\pm0.08$ & $15.34\pm0.15$ & 27 May 1999 & Keck NIRC & no\tablenotemark{b}\nl
2MASSW J2142333$-$035556 & B & $15.86\pm0.06$ & $15.69\pm0.14$ & $15.73\pm0.24$ & 27 May 1999 & Keck NIRC & no\tablenotemark{b}\nl

\tablenotetext{a}{Source designations for 2MASS discoveries are given
as ``2MASSx Jhhmmss[.]s$\pm$ddmmss'', where the ``x'' prefix varies depending upon
which catalog the object originates, in this case ``I'' for 1999 Spring Release
Data and ``W'' for the survey's working database.  The suffix conforms to IAU
nomenclature convention and is the sexigesimal R.A. and decl. at J2000 equinox.} &
\tablenotetext{b}{Probable uncatalogued asteroid.}
\tablenotetext{c}{Not detected at K$_s$ band; given magnitude is a 95\% confidence
magnitude lower (bright) limit based on the background flux.}
\tablenotetext{d}{This object was previously discovered by the Sloan Digitized Sky
Survey (Tsvetanov et al. 1999).}
\enddata
\end{deluxetable}

\clearpage

\begin{deluxetable}{lcccccc}
\scriptsize
\tablenum{2}
\tablewidth{0pt}
\tablecaption{Confirmed 2MASS T dwarfs. \label{tbl-2}}
\tablehead{
\colhead{Object} &
\colhead{J} &
\colhead{J-H} &
\colhead{H-K$_s$} &
\colhead{J-K$_s$} &
\colhead{est. distance (pc)\tablenotemark{a}} \nl 
\colhead{(1)} &
\colhead{(2)} & 
\colhead{(3)} & 
\colhead{(4)} & 
\colhead{(5)} & 
\colhead{(6)} 
}
\startdata
2MASSW J1217111$-$031113 & $15.85\pm0.07$ & {\phs}$0.06\pm0.14$ & $< -0.12$ & $< -0.06$ & 14\nl
2MASSW J1346464$-$003150\tablenotemark{b} & $15.86\pm0.08$ & $-0.19\pm0.22$ & $< 0.30$ & $< 0.11$ & 14\nl 
2MASSW J1225543$-$273947 & $15.23\pm0.05$ & {\phs}$0.13\pm0.10$ & {\phs}$0.04\pm0.17$ & {\phs}$0.17\pm0.16$ & 10\nl
2MASSI J1047539$+$212423 & $15.82\pm0.06$ & {\phs}$0.03\pm0.13$ & $< -0.50$ & $< -0.47$ & 13\nl
2MASSW J1237392$+$652615 & $15.90\pm0.06$ & {\phs}$0.03\pm0.14$ & $< -0.03$ & $< 0.00$ & 14\nl

\tablenotetext{a}{Distance estimates assuming Gl 229B-like radii and 
J-band bolometric corrections, and T$_{eff}$ = 1000 K.}
\tablenotetext{b}{This object was previously discovered by the Sloan Digitized Sky
Survey (Tsvetanov et al. 1999).}
\enddata
\end{deluxetable}

\clearpage

\plotone{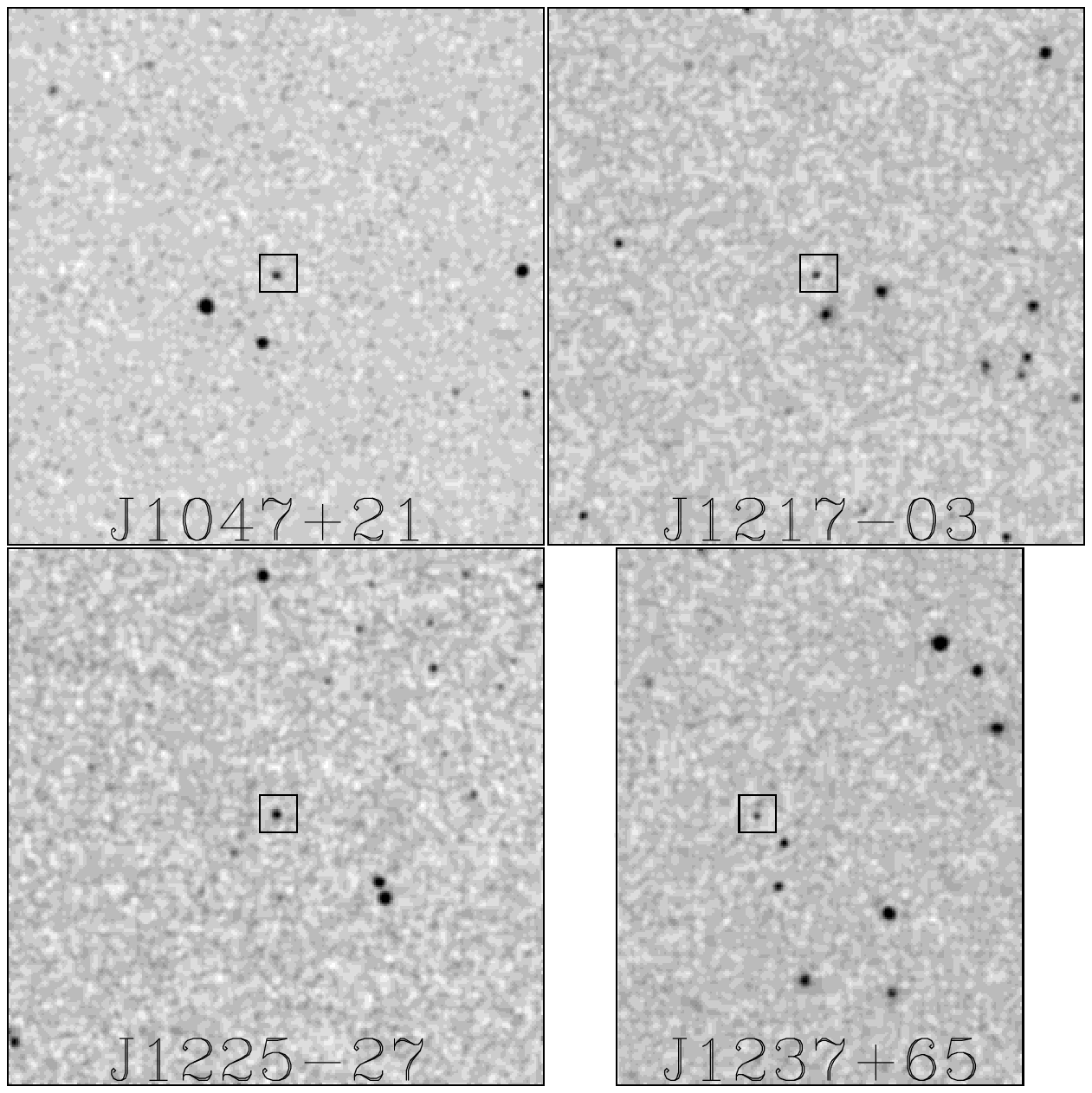}

\figcaption[fig1a.ps]{J-band images of the four new 2MASS T dwarfs:
2MASS J1047+21, J1217-03, J1225-27, and J1237+65.  
Fields are 5$\arcmin$ x 5$\arcmin$ (except for 2MASS J1237+65 which
is 3.8$\arcmin$ x 5$\arcmin$) with north up and east to the left.  A 20$\arcsec$ x 20$\arcsec$
box is drawn around each T dwarf. \label{fig-1}}

\clearpage


\plotone{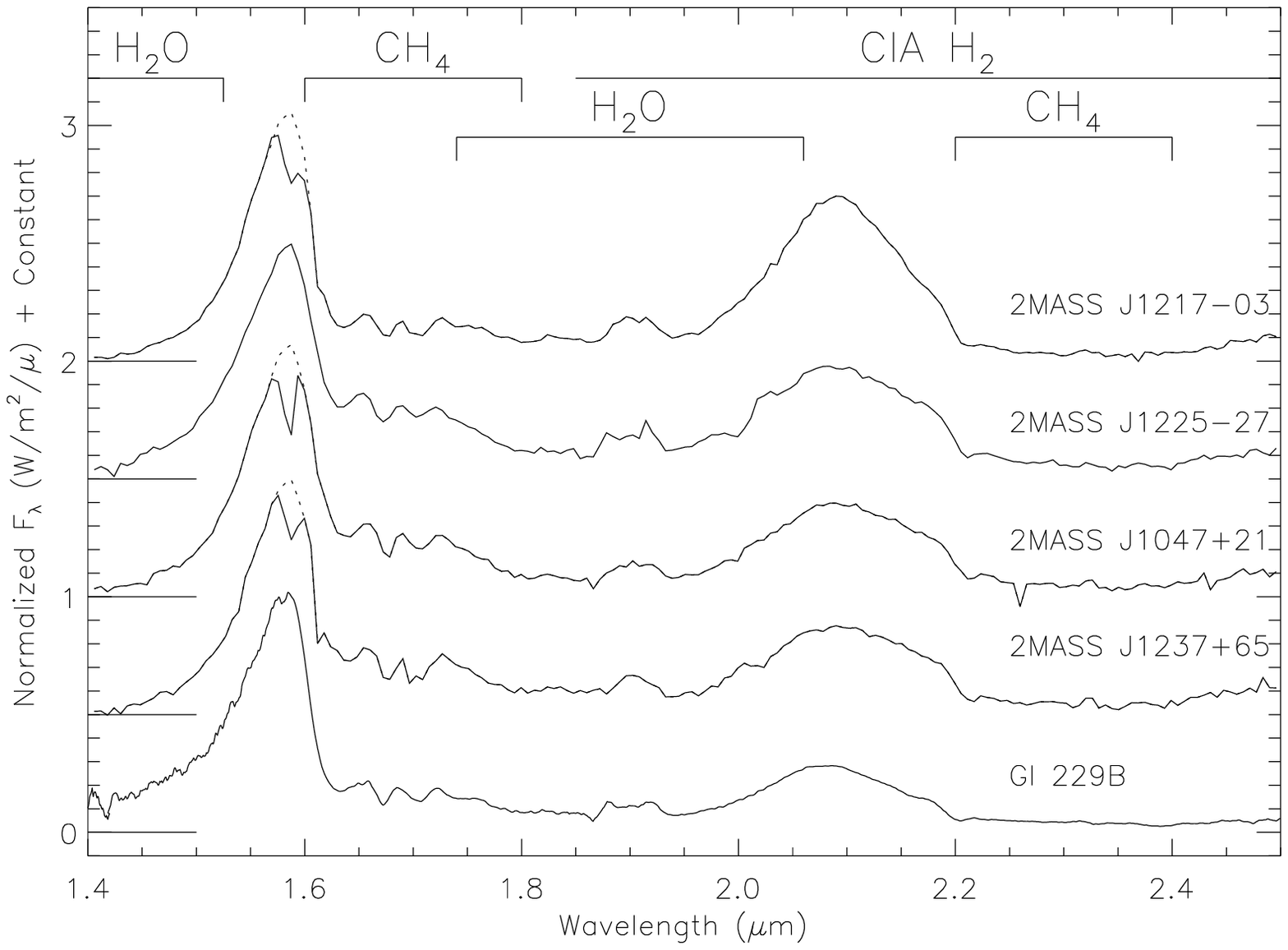}

\figcaption[fig2a.ps]{Near-infrared spectra (1.4 to 2.4 $\micron$) of the four
new 2MASS T dwarfs, taken with the Keck I NIRC grism, along with Gl 229B spectral data
from Leggett et al. (1999).  The Gl 229B data has been smoothed to match
the resolution of our NIRC observations.  Objects are displayed from
top to bottom in a preliminary temperature sequence: 2MASS J1217-03,
J1225-27, J1047+21, J1237+65, and Gl 229B.  Absorption bands for
H$_2$, H$_2$O, and CH$_4$ are indicated.  The ``absorption features''
seen at 1.58 $\micron$ for all but 2MASS J1225-27 are due to saturation, and the
dotted lines are extrapolations of the data in this region to fit to the 2MASS J1225-27
H-band peak.  Spectra are normalized to their H-band peak.
\label{fig-2}}


\end{document}